\documentclass[aps,twocolumn]{revtex4}

\usepackage{amsmath,amssymb,graphicx}
\usepackage{algorithmic}
\usepackage{enumerate}
\usepackage{color}

\begin{document}
\title{Measuring free energy in parallel tempering Monte Carlo}
\author{Wenlong Wang}
\email{wenlong@physics.umass.edu}
\affiliation{Department of Physics, University of Massachusetts,
Amherst, Massachusetts 01003 USA}

\begin{abstract}
An efficient approach of measuring the absolute free energy in parallel tempering Monte Carlo using the exponential averaging method is discussed and the results are compared with those of population annealing Monte Carlo using the three-dimensional Edwards-Anderson Ising spin glass model as benchmark tests. Numerical results show that parallel tempering, even though uses a much less number of temperatures than population annealing, can nevertheless equally efficiently measure the absolute free energy by simulating each temperature for longer times.
\end{abstract}

\maketitle

\section{INTRODUCTION}
Measuring free energy is difficult in Monte Carlo simulations, but nevertheless can be very helpful if available. More phenomena and finite size effects can be analyzed using free energy. This is not only for the sake of free energy, but also as a consequence entropy can be obtained indirectly since energy can be readily measured in Monte Carlo simulations. The Bennett acceptance ratio method \cite{Bennett} is mainly used to compute the free energy difference of two different systems but with the same state space at the same temperature. To access free energy differences of the same system at two different temperatures, exponential averaging \cite{freeenergy} and thermodynamic integration \cite{chaos1,TI} are often used. Thermodynamic integration has the drawback of integration errors and therefore needs data at many temperatures to be accurate. When data is available at only few temperatures interpolation techniques are often needed. This is especially a problem when the specific heat shows complicated multi-peak structures like in spin glass \cite{TC2015} and protein folding \cite{PF} problems due to various reasons like phase transitions and temperature chaos. The exponential averaging method however doesn't have this problem and therefore is still beneficial and suitable for such models.

For systems with complicated free energy landscapes like spin glasses, a simple Monte Carlo method is not sufficient. Two algorithms that both have a hierarchical structure are shown to be efficient: parallel tempering \cite{ptmc1,ptmc2,ptmc3} and population annealing \cite{F,A}. It has been known for more than a decade that population annealing can accurately estimate the absolute free energy \cite{A} using the exponential averaging method. But slightly negative view has been held for parallel tempering due to mainly two reasons: the first is that there are too few temperatures, therefore the measurement would not be very accurate while the other is that the highest temperature is usually finite for parallel tempering while in population annealing this can be naturally chosen as infinity. The second problem can be trivially solved by adding a high temperature stage before the parallel tempering stage using simple single temperature Monte Carlo since thermal equilibration is fast at high temperatures. For the first problem, here we show numerically that the larger temperature steps in parallel tempering compared with population annealing are compensated by simulating each temperature for longer times, therefore, permitting equally accurate measurements of free energy in parallel tempering. Parallel tempering however does suffer some minor technical problems, but nevertheless can be readily solved to get access to free energy.

The work is organized as follows: Section~\ref{model} introduces the Edwards-Anderson model, the free energy estimator using the exponential averaging method and the two-stage free energy measurement algorithm in parallel tempering. The results of parallel tempering and population annealing are compared in Sec.~\ref{result} and the conclusions are stated in Sec.~\ref{conclusion}.

\section{MODELS AND METHODS}
\label{model}
\subsection{The Edwards-Anderson model}
The Edwards-Anderson (EA) Hamiltonian is defined as
\begin{equation}
\mathcal{H}=-\sum\limits_{\langle ij \rangle} J_{ij} s_i s_j,
\end{equation}
where ${s_i}$ are the spin degrees of freedom defined on a three-dimensional cubic lattice with $s_i=\pm1$. The sum over $\langle ij \rangle$ means sum over the nearest neighbour sites. $J_{ij}$ is the coupling between spin $s_i$ and $s_j$ and is independently chosen from the standard Gaussian distribution with mean zero and variance one. I will refer to each disorder realization as a sample. Periodic boundary conditions are used in the simulations of this work.

\subsection{The free energy estimator}
It has been known that population annealing (PA) can \textit{naturally} estimate the absolute free energy of the EA model \cite{A,B}. The method used, exponential averaging, is nevertheless general. The ratio of $Z(\beta^{'})$ and $Z(\beta)$, which are the partition functions at inverse temperatures $\beta^{'}$ and $\beta$ respectively are
\begin{align}
\dfrac{Z(\beta^{'})}{Z(\beta)}&=\dfrac{\sum\limits_{s} e^{-\beta^{'} E_s}}{Z(\beta)} \\
   &=\sum\limits_{s} e^{-(\beta^{'}-\beta)E_s} \left( \dfrac{e^{-\beta E_{s}}}{Z(\beta)} \right) \\
   &=\langle e^{-(\beta^{'}-\beta) E_{s}}\rangle_{\beta} \\
   &\approx \dfrac{1}{N_0} \sum\limits_{i=1}^{N_0} e^{-(\beta^{'}-\beta) E_i} \\
   &=Q(\beta,\beta^{'}),
\end{align}
where the sum over $s$ is the sum over all states and the sum over $i$ is the sum over measured equilibrium states in a Monte Carlo run. $E$ is the energy of a micro-state, $N_0$ the number of measurements and $\langle...\rangle$ means a thermal average. Take the natural logarithm of both sides
\begin{eqnarray}
\ln Z(\beta^{'})-\ln Z(\beta)=\ln Q(\beta,\beta^{'}) \\
-\beta^{'} F(\beta^{'})=-\beta F(\beta)+\ln Q(\beta,\beta^{'}),
\end{eqnarray}
where $F$ is the free energy. Note that by sampling equilibrium states at a temperature, one can integrate the free energy between the temperature and a nearby temperature, usually chosen as a lower one. The initial condition of $\beta F$ is known at $\beta=0$ theoretically. At $\beta=0$, $\ln Z =\ln \Omega$, where $\Omega$ is the total number of micro-states and $\Omega=2^N$ with $N=L^3$ the total number of spins. If we order a set of temperatures as $\beta_0<\beta_1<...<\beta_k$ and $\beta_0=0$, then we have
\begin{equation}
-\beta_i F(\beta_i)=\sum\limits_{j=0}^{j=i-1} \ln Q(\beta_{j+1},\beta_j)+\ln \Omega.
\end{equation}

\subsection{Measuring free energy in parallel tempering}
Since parallel tempering (PT) doesn't usually work from $\beta=0$, we need to modify the parallel tempering algorithm somewhat to get the absolute free energy. One simple way is to implement the simulation in two stages with a high temperature stage using simple single temperature Monte Carlo and a regular parallel tempering Monte Carlo stage. The two stages together will cover from $\beta=0$ to a low temperature of interest.
\begin{itemize}
\item \textit{Simple Monte Carlo stage} \\
Suppose the regular parallel tempering Monte Carlo works between $T_{\rm{min}}$ and $T_{\rm{max}}$, the minimum and maximum temperatures, respectively. The first stage is to use simple single temperature Monte Carlo to simulate the system and work from $\beta=0$ to $1/T_{\rm{max}}$, not including $1/T_{\rm{max}}$. Then the free energy can be integrated from $\beta=0$ to $1/T_{\rm{max}}$, including $1/T_{\rm{max}}$.
\item \textit{Parallel tempering Monte Carlo stage} \\
Simulate the system using the regular parallel tempering Monte Carlo between $T_{\rm{min}}$ and $T_{\rm{max}}$. Then the free energy can be further integrated to the low temperature spin glass phase.
\end{itemize}

In my implementation, I used $N_{\beta}$ temperatures evenly distributed in $\beta$ for the first stage while used $N_T$ temperatures evenly distributed in $T$ for the second stage. A uniform distribution in $\beta$ for the first stage is easier to work with because of the infinite temperature. For the second stage, a uniform distribution in $T$ can yield better performance of parallel tempering since the swap probabilities are more constant at different temperatures. Also, no Monte Carlo sweeps were done at $\beta=0$, it is sufficient to generate random states and make measurements like that of population annealing. In this work, the amount of work is counted in terms of Monte Carlo sweeps and one Monte Carlo sweep is a sequential update of all the spins for one replica at one temperature once. The same number of sweeps were done for both the thermal equilibration run and the data collection run.

It is important to stress here that \textit{the only difference of this algorithm compared with the regular parallel tempering algorithm is the additional simple Monte Carlo stage.} Since the free energy landscape is not rough at high temperatures, the autocorrelation time in sweeps therefore doesn't depend on the energy landscape and system size. It is not necessary to do as many sweeps as in the second stage, where the free energy landscape is rough. Therefore when the algorithm is well optimized, most of the work would be spent in the second stage and the overhead would be small when adding the first stage. In fact, the work spend in the first stage compared with the second stage should vanish in the thermodynamic limit. In addition, due to the similarity of this algorithm with the regular parallel tempering algorithm, it is straightforward to modify a regular parallel tempering code to collect the free energy data.

Finally, I want to point out that there is a technical issue of overflow when computing $Q$ for large systems if $N_T$ is small or there are too many data collection steps for parallel tempering. The later problem can be easily solved since data collection after each Monte Carlo sweep is neither necessary nor desired. For the former problem, one can subtract a low energy from the energy of a state when computing the exponential term and then add it back when integrating $-\beta F$. Another method is averaging $Q$ on the fly. One can also use reasonably more temperatures, but this requires more computational work and may also increase the round trip time. Note that this problem could appear as well for population annealing. However, since population annealing naturally uses a lot more temperatures and the temperatures are usually evenly distributed in $\beta$, no such problem was encountered in our studies of the population annealing algorithm at least up to size $L=14$ down to $\beta=3$ \cite{pamc}.

\section{RESULTS}
\label{result}
The free energy of  each sample was measured using both the parallel tempering and population annealing algorithms from infinite temperature to low temperatures deep in the spin glass phase. The parallel tempering results are compared with the production run of population annealing \cite{TBC}. The reference simulation parameters of population annealing are summarized in Table~\ref{para1} while the simulation parameters of parallel tempering are summarized in Table~\ref{para2}. Since the free energy results of the two algorithms agree to a very high degree of precision, I have therefore studied a random subset of samples of \cite{TBC}, 100 samples per system size. This is sufficient to show the equally efficiency of parallel tempering in measuring free energy compared with population annealing. In the following, I will first make a detailed comparison for a single hard sample and then a large scale comparison for the two algorithms.

\begin{table}
\caption{
Parameters of the reference runs of PA \cite{TBC} for different system sizes $L$ with periodic boundary conditions. $R$ represents the number of replicas, $1/\beta_0$ is the lowest temperature
simulated, $N_T$ the number of temperatures used in the annealing
schedule, $N_S$ the number of sweeps per temperature and $M$ the number
of samples.
\label{para1}
}
\begin{tabular*}{\columnwidth}{@{\extracolsep{\fill}} l c c c c c l l}
\hline
\hline
$L$  & $R$ & $1/\beta_0$ & $N_T$ & $N_S$ & $M$ \\
\hline
$4$  & $5\,10^4$ & $0.2$     & $101$ & $10$  & $100$ \\
$6$  & $2\,10^5$ & $0.2$     & $101$ & $10$  & $100$ \\
$8$  & $5\,10^5$ & $0.2$     & $201$ & $10$  & $100$ \\
$10$ & $   10^6$ & $0.2$     & $301$ & $10$  & $100$ \\
\hline
\hline
\end{tabular*}
\end{table}

\begin{table}
\caption{
Parameters of PT for different system sizes $L$ with periodic boundary conditions. $N_{\beta}$ represents the number of temperatures in the simple Monte Carlo stage, $N_T$ the number of temperatures in the PT stage, $T_{\rm{min}}$ the lowest temperature
simulated, $T_{\rm{max}}$ the highest temperature simulated in the PT stage, $N_S$ the number of sweeps per temperature and $M$ the number
of samples.
\label{para2}
}
\begin{tabular*}{\columnwidth}{@{\extracolsep{\fill}} l c c c c c l l}
\hline
\hline
$L$  & $N_{\beta}$ & $N_T$ &$T_{\rm{min}}$ &$T_{\rm{max}}$ & $N_S$ & $M$ \\
\hline
$4$  & $5$ &10 & $0.2$     & $2.0$ & $5\,10^5$  & $100$ \\
$6$  & $5$ &20 & $0.2$     & $2.0$ & $10^6$  & $100$ \\
$8$  & $10$ &30 & $0.2$     & $2.0$ & $5\,10^6$  & $100$ \\
$10$ & $10$ &40 & $0.2$     & $2.0$ & $5\,10^7$  & $100$ \\
\hline
\hline
\end{tabular*}
\end{table}

\subsection{Detailed comparison of a single hard sample}

In this section, I will do a detailed comparison of the efficiency of parallel tempering and population annealing in measuring the free energy using the hardest sample out of about 5000 samples of $L=8$ \cite{TBC}. I will first focus on the dimensionless quantity of $-\beta F$ at a low temperature $T=0.2$ and study how the mean and the errorbar of the systematic error of $-\beta F$ change as a function of the amount of work $W$. Then I will study how the relative errors of the estimated free energy evolves as a function of temperature at a fixed amount of work. The amount of work is counted as the total number of sweeps performed in the simulation.

\begin{figure}[htb]
\begin{center}
\includegraphics[scale=0.68]{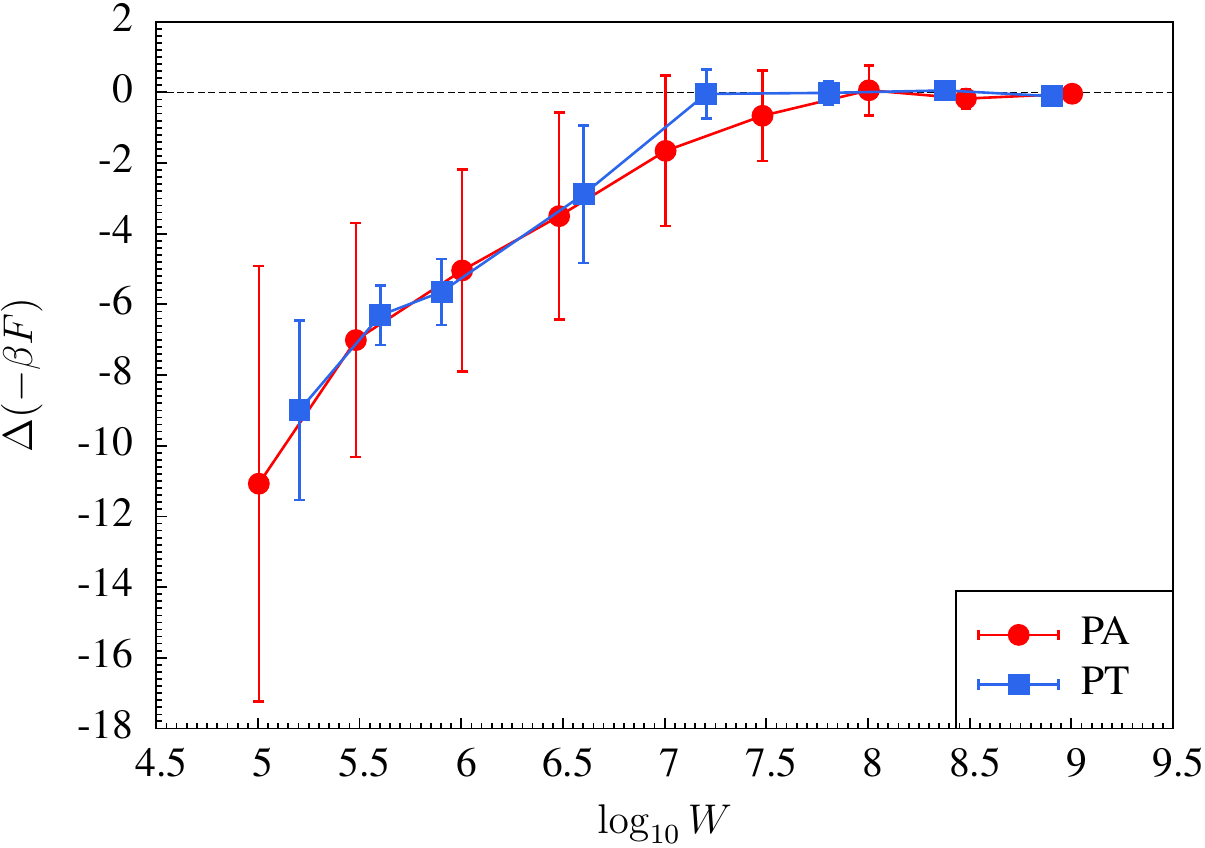}
\caption{Linear-log plot of the systematic error $\Delta(-\beta F)$ as a function of the amount of work for a hard sample of $L=8$ at $T=0.2$. The errorbar is the the standard deviation of the $-\beta F$ distribution computed from multiple runs, not the errorbar of the sample mean of $-\beta F$. The exact value is taken from a large scale simulation using population annealing \cite{TBC}.}
\label{papt}
\end{center}
\end{figure}

The systematic error $\Delta(-\beta F)$ is defined as the differences of the estimated $\beta F$ and the exact $-\beta F$ i.e. $\Delta(-\beta F)=-\beta F +\beta F_{\rm{exact}}$. The exact $-\beta F$ is \textit{estimated} %to be $4457.610$
using the reference runs of population annealing  \cite{TBC}. The population annealing comparison data with different amount of work is taken from a previous study of finding ground states \cite{2014GS}. The work is varied by changing the number of replicas while holding $N_T=101$ and $N_S=10$ constant. The parallel tempering data was done using the parameters of Tabel~\ref{para2} but varying $N_S$. The result is shown in Fig.~\ref{papt}. One thing we can learn from this study is that the accuracy is about the same considering neither algorithms is very carefully optimized. Furthermore, both algorithms underestimate the value of $-\beta F$ when the amount of work is too small i.e. the sample is not in thermal equilibrium. This can be understood as those runs are so small so that some low energy states were probably not properly found so that $Q$ was underestimated, and therefore also $-\beta F$.

\begin{figure}[htb]
\begin{center}
\includegraphics[scale=0.78]{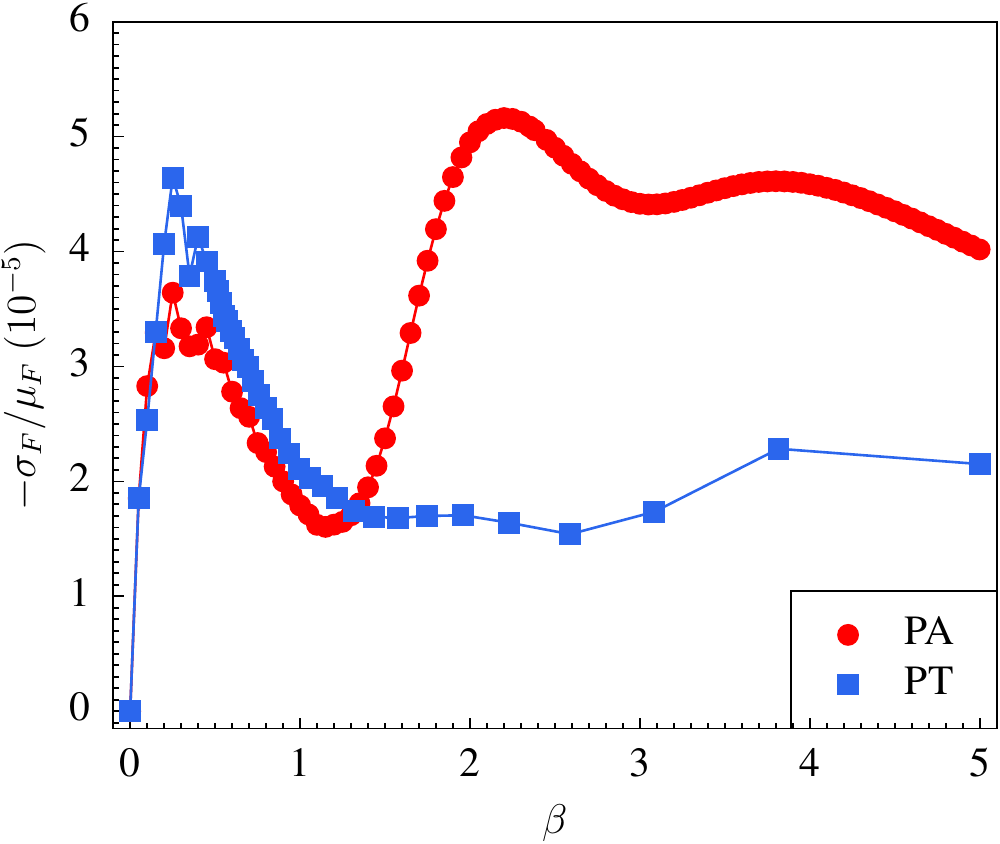}
\caption{Evolution of the relative error of the estimated free energy $-\sigma_F/\mu_F$ at a function of the inverse temperature $\beta$ for a hard sample of $L=8$. The data is the same as the largest run of Fig.~\ref{papt} for both PA and PT. Note that the magnitude of the relative error does not grow as temperature is lowered.}
\label{ft}
\end{center}
\end{figure}

It is also important and interesting to study how the relative error of the estimated free energy behaves as temperature is lowered due to the integration nature of the method. The relative error is defined as minus the ratio of the standard deviation $\sigma_F$ and mean $\mu_F$ of the estimated free energy. The minus sign is to ensure the defined relative error is positive. The relative error is measured via multiple runs in practice. Figure~\ref{ft} shows this quantity as a function of inverse temperature $\beta$ for both PA and PT for the largest run of Fig.~\ref{papt}. Note that there is no trend that the relative error grows as temperature is lowered, showing the effectiveness of both algorithms. In addition, the magnitude of the relative errors is also about the same.

\subsection{A large scale comparison}
\label{lsc}
\begin{figure}[htb]
\begin{center}
\includegraphics[scale=0.68]{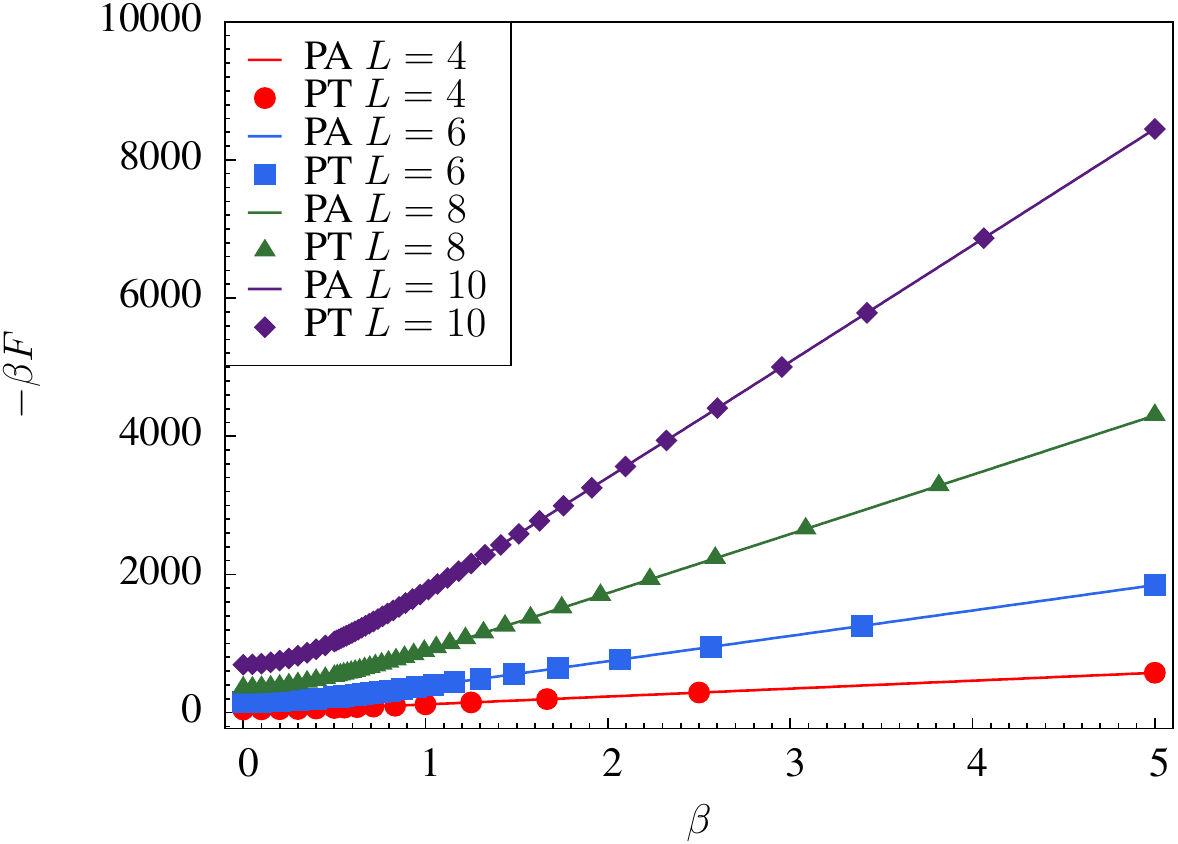}
\caption{Comparison of $-\beta F$ for a typical sample of system sizes $L=4,6,8$ and 10 in a wide range of temperatures. The PT data falls right on top of the PA curve, showing the effectiveness of PT in measuring free energy.}
\label{FEs}
\end{center}
\end{figure}

Before showing the large scale comparison, I would like to show a comparison of the result of $-\beta F$ for a typical sample of each system size in the whole range of temperatures studied. A typical result is shown in Fig.~\ref{FEs}. Note that the parallel tempering data falls right on top of the population annealing curve, showing the effectiveness of parallel tempering in measuring free energy in a wide range of temperatures.

\begin{figure}[htb]
\begin{center}
\includegraphics[scale=0.9]{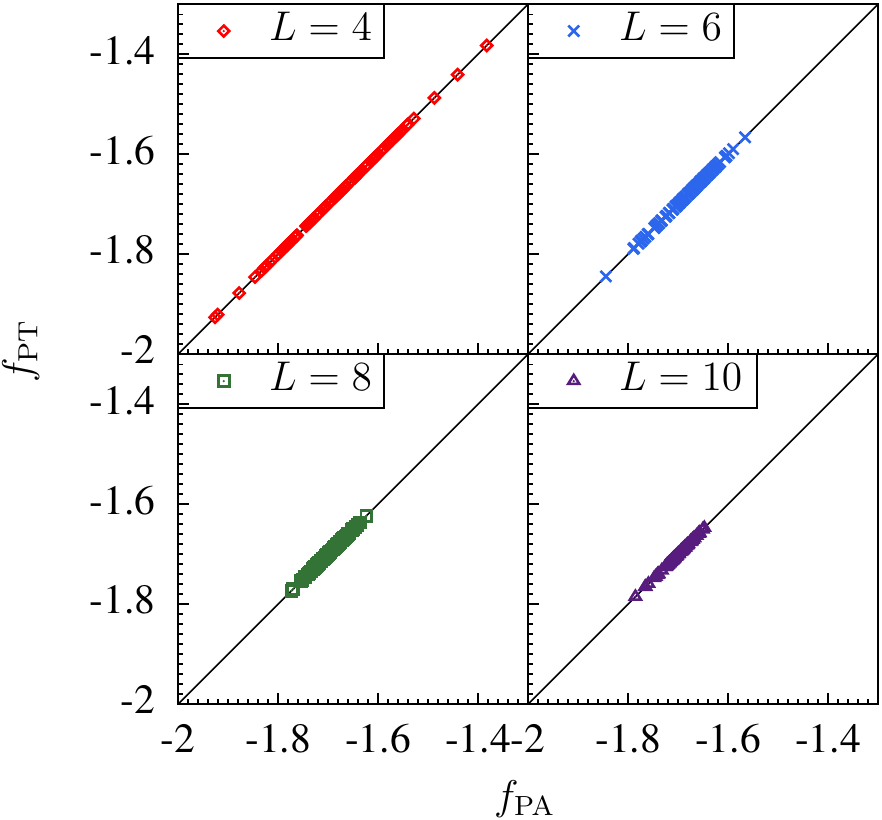}
\caption{Scatter plot of the free energy per spin $f$ of PA and PT of system sizes $L=4,6,8$ and 10 at $T=0.2$. Each point represents a sample.}
\label{FE}
\end{center}
\end{figure}

\begin{figure}[htb]
\begin{center}
\includegraphics[scale=0.78]{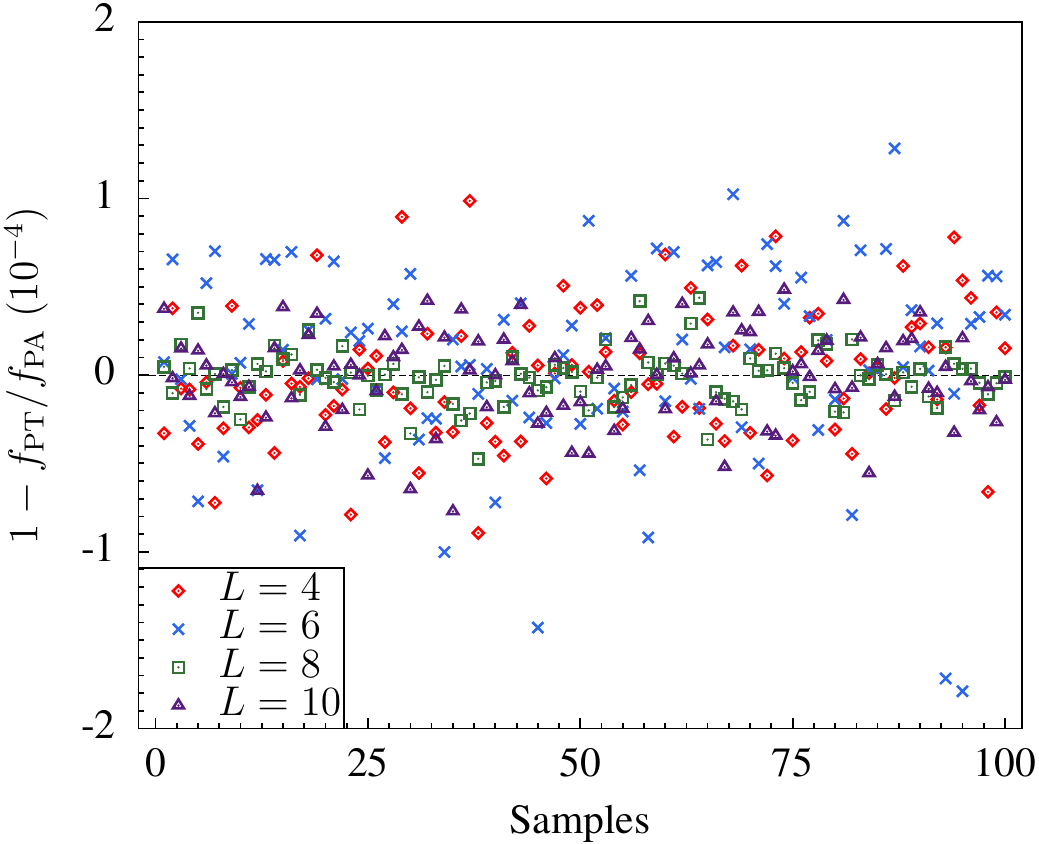}
\caption{The same data as in Fig.~\ref{FE}, but plotting the relative error $1-f_{\rm{PT}}/f_{\rm{PA}}$ of the free energy per spin of PT against PA at $T=0.2$. Each point represents a sample.}
\label{FERD}
\end{center}
\end{figure}

To make a comparison of more samples, a scatter plot of the free energy per spin $f$ at the lowest simulation temperature $T=0.2$ of PA and PT is shown in Fig.~\ref{FE}. Note that the statistical error compared with the absolute value is too small to be seen in this plot, it is therefore interesting to look at the relative error $1-f_{\rm{PT}}/f_{\rm{PA}}$ of the free energy per spin of parallel tempering against population annealing. The relative error is shown in Fig.~\ref{FERD}. The errors are well bounded within the accuracy of about $10^{-4}$ and are well scattered around zero suggesting the nature of the errors is essentially statistical and again showing PT is as efficient as PA in accurately measuring the absolute free energy.

\section{CONCLUSION}
\label{conclusion}
In this paper, I showed an efficient approach to measure the absolute free energy in parallel tempering using the exponential averaging method. The algorithm is tested using the three-dimensional EA model and the results of parallel tempering agree very well with those of population annealing, showing parallel tempering can equally efficiently measure free energy as population annealing. There are technical issues with parallel tempering but nevertheless can be readily solved. The fact that the reweighting technique works so well suggests that many quantities including energy and free energy can be accurately estimated using the reweighting technique at many temperatures with little overhead in parallel tempering without using interpolation techniques. Finally, I hope that this simple and direct access to free energy in parallel tempering can make the spin glass and other relevant research fields potentially richer.

\section*{Acknowledgments}
I gratefully acknowledge support from NSF (Grant No.~DMR-1208046). I thank Helmut Katzgraber for kindly allowing me to use part of the reference population annealing data. I thanks Jon Machta and Helmut Katzgraber for helpful discussions and suggestions, and also for the careful reading of the manuscript.

% use  \cite{Aa10} for references;
%\bibliographystyle{plain}
%\bibliographystyle{apsrevtitle}
%\bibliography{references,refs}

\end{document}